\documentclass[sigconf]{acmart}
\renewcommand\footnotetextcopyrightpermission[1]{} 
\newcommand{\pname}{SIMDive}
\usepackage{algorithmic}
\usepackage{graphicx}
\usepackage{textcomp} 
\usepackage{xcolor}
\usepackage{url} 
\usepackage{pifont}
\newcommand{\cmark}{\ding{51}}%
\newcommand{\xmark}{\ding{55}}%

\newcommand{\mysize}{\fontsize{10pt}{12pt}\selectfont} 
\newcommand{\HUGER}{\fontsize{17pt}{20pt}\selectfont} 

\usepackage{booktabs, caption, makecell}

\usepackage{threeparttable}     

\usepackage{tabu}

\usepackage{leftidx}

\usepackage{rotating} 
\usepackage{multirow}
\usepackage{subfig} 
\usepackage{soul} 

\usepackage{enumitem}
\usepackage{mathtools} 
\usepackage{textcomp} 
\usepackage{siunitx} 
\usepackage{amsmath}
\usepackage[utf8]{inputenc} 
\definecolor{salim}{rgb}{0.7,0.0,0}
\definecolor{forestgreen}{rgb}{0.13, 0.55, 0.13}

\usepackage{xcolor,colortbl}
\usepackage[font=footnotesize]{caption}

\definecolor{lightgray}{gray}{0.75}
\definecolor{LightCyan}{rgb}{0.88,1,1}
\definecolor{green}{rgb}{0.45,0.85,0.65}
\definecolor{sam_color}{rgb}{1.0,0.3,0.2}
\newcolumntype{a}{>{\columncolor{lightgray}}c}

\usepackage{tabularx}

\usepackage{tikz}

\newcolumntype{L}{>{\centering\arraybackslash}m{1.2cm}}
\usetikzlibrary{arrows,fit,positioning}

\usetikzlibrary{calc}
\usetikzlibrary{decorations.pathmorphing}
\newcolumntype{C}[1]{>{\centering\let\newline\\\arraybackslash\hspace{0pt}}m{#1}}
\usepackage{ragged2e}
\newcolumntype{x}[1]{>{\centering\hspace{0pt}\arraybackslash}p{#1}}

\newcommand{\figbox}[1]{%
  \fbox{%
    \vbox to 1in{%
    \vfil
    \hbox to 2in{%
      \hfil
      #1%
      \hfil}%
    \vfil}}}

\newcommand{\shorteq}{%
  \settowidth{\@tempdima}{-}
  \resizebox{\@tempdima}{\height}{=}%
  }

\emergencystretch=\maxdimen
\usepackage[all]{background}
\usepackage{lipsum}
\SetBgContents{\textcolor{forestgreen}{This paper is accepted to be published in ACM Great Lakes Symposium on VLSI (GLSVLSI) 2020 (author-ready version) }}
\SetBgPosition{-1.1cm,-9.0cm}
\SetBgOpacity{1.0}
\SetBgAngle{90.0}
\SetBgScale{1.3}

\begin{document}
\fancyhead{}

\title{SIMDive: Approximate \underline{SI}MD Soft \underline{M}ultiplier-\underline{Div}ider for FPGAs with Tunabl\underline{e} Accuracy}


\author{Zahra Ebrahimi, Salim Ullah, Akash Kumar} 
    \affiliation{ 
      \city{Technische Universit\"{a}t Dresden} 
      \state{Germany} 
    \postcode{Corresponding\;Author's Email: zahra.ebrahimi\_mamaghani@tu-dresden.de}
    }
    \email{zahra.ebrahimi\_mamaghani@tu-dresden.de}

\begin{abstract}

The ever-increasing quest for data-level parallelism and variable precision in ubiquitous multimedia and \emph{Deep Neural Network} (DNN) applications
has motivated the use of\! \emph{Single Instruction, Multiple Data} (SIMD) architectures. To alleviate energy as their main resource constraint, approximate computing has re-emerged,
albeit mainly specialized for their \emph{Application-Specific Integrated Circuit} (ASIC) implementations.
This paper, presents for the first time, an SIMD architecture based on novel multiplier and divider with tunable accuracy, targeted for \emph{Field-Programmable Gate Arrays} (FPGAs). The proposed
hybrid architecture implements Mitchell’s algorithms and supports precision variability from 8 to 32 bits. Experimental results obtained from Vivado, multimedia and DNN applications indicate superiority of proposed architecture (both SISD and SIMD) over accurate and state-of-the-art approximate counterparts. In particular, the proposed SISD divider outperforms the accurate \emph{Intellectual Property} (IP) divider provided by Xilinx with \texttildelow4{$\times$} higher speed and 4.6{$\times$}\! less energy and tolerating only\,$<$0.8\% error. Moreover, the proposed SIMD multiplier-divider supersede accurate SIMD multiplier by
achieving up to 26\%, 45\%, 36\%, 
and 56\% improvement in area, throughput, power, 
and energy, respectively. 

	
 \vspace{-0cm}
\end{abstract}



\maketitle

\section{ Introduction} \label{introduction}
The computationally-intensive nature of upcoming \emph{Internet of Things} (IoT) and multimedia, has dictated the demand to feature energy-efficient, multi-precision \emph{Single Instruction, Multiple Data} (SIMD) architectures. Approximate computing paradigm has shown to serve as a viable energy-efficient solution for these applications after the strives to prevent breakdown of Moore's law with the cease of Dennard scaling era \cite{jain2018compensated}. This technique has also become pronounced in
machine learning domain, as it can trade stringent resource budget with a tolerable quality relaxation.\\
Multipliers and dividers are the most highly-used/resource-hungry arithmetic units in the kernel of these applications (dominating 99\% of computational energy \cite{jain2018compensated}) which carves out a prominent niche for their approximation. In particular, long latency of divider limits the overall speed of applications. Approximation of division has recently gained attention as, although less frequent, this operation is still inevitable in these applications. For instance, it is used within K-means in unsupervised clustering, Discrete Cosine Transform in JPEG compression and AlexNet Convolutional Neural Network (up to half a million times in three layers) \cite{krizhevsky2012imagenet}. However, the required numerical-precision for these operations is not fixed among all or even within an application (e.g., not only the precision varies during learning and retrieval phase, but also among layers of a neural network \cite{judd2016stripes}).
\begin{table*}[t]
\centering
\caption{{Summary of SISD (approximate) and SIMD (accurate/approximate) multipliers and dividers in the literature}}
\vspace{-0.3 cm}
\label{table:related}
\resizebox{1.0\textwidth}{!}{
\begin{tabular}{|c|c|c|c|c|c|c|c|c|}
\hline
\textbf{SIMD} & \textbf{Approx} & \textbf{Mul/Div} &
\textbf{Description of work}                                                                                                         & \textbf{Platform} & \textbf{Improvement}                     & \textbf{Avg\,Rel\,Error\,(\%)}\\ \hline
\xmark & \cmark  & \cmark \,\! \slash \,\! \xmark                                          & 4x4 and 8x8 with approximate partial products        \cite {ullah2018smapproxlib}                                                       & FPGA     & \{Area, power, latency\}+  & 1.6         \\ \hline
\xmark & \cmark   & \cmark \,\! \slash \,\! \xmark                                               & 4-, 8-, and 16-bit  with approximate PPs using 4x2 instances   \cite {ullah2018area}                                            & FPGA     & \{Area, energy\}+ & 0.3        \\ \hline
\xmark & \cmark  & \cmark \,\! \slash \,\! \xmark              &   Improving accuracy of Mitchell's logarithmic multiplication \cite{saadat2018minimally} & ASIC     & \{Area, power\}++              & 2.7         \\ \hline
\xmark & \cmark  & \xmark \,\! \slash \,\! \cmark               &  Improving accuracy of Mitchell's division (same approach in \cite{saadat2018minimally} is applied for division) \cite {saadat2019approximate} & ASIC     & \{Area, power\}++, Delay+++              & 2.9         \\ \hline
\xmark & \cmark & \xmark \,\! \slash \,\! \cmark              & MSB division based on position of leading-one (\cite{jiang2018adaptive}, \cite{hashemi2016low})     & ASIC     & \{Delay, power, energy\}++              & 0.8 to 3, 13.4         \\ \hline
\xmark & \cmark & \xmark \,\! \slash \,\! \cmark             & Multiplying truncated/rounded dividend with reciprocal of divisor: \cite{vahdat2017truncapp}, \cite{Vaeztourshizi:2018:EYH:3218603.3218650}, 
\cite{Behroozi:2019:SSA:3287624.3287668}     & ASIC     & \{Area, power, delay\}+++              & 4.2, 2.8, 2.4, 4.9        \\ \hline
\xmark & \cmark & \xmark \,\! \slash \,\! \cmark              & Inexact subtractors in array divider \cite{8464807}, Approximating FP mantissa division to subtraction \cite{8715112}      & ASIC     & \{Area, power, delay\}+              & 1.8, 0.6 for FP         \\ \hline 
\cmark & \xmark  & \cmark \,\! \slash \,\! \xmark                                              &Accurate variable-precision multiplier (8x8 to 32x32) \cite {perri2004variable}                        & FPGA     & Throughput ++        & Accurate          \\ \hline
\cmark & \xmark   & \cmark \,\! \slash \,\! \xmark           & 8- to 32-bit Add/Sub/Mul for FPGA-based multimedia processors   \cite {lanuzza2005low}, \cite {purohit2008power}                        & FPGA     & Throughput ++          & Accurate          \\ \hline
\cmark & \cmark    & \cmark \,\! \slash \,\! \xmark               & Variable-precision multiplier based on 8-bit truncated instances \cite {osorio2019truncated}         & ASIC     & \{Area, Energy\} +             & 1.2   \\ \hline
\rowcolor{lightgray}
\cmark & \cmark  & \cmark \,\! \slash \,\! \cmark                      & First approximate hybrid multiplier/divider (SISD and SIMD) with tunable accuracy
& FPGA     & Area ++, \{Throughput, energy\}+++         & 0.7   \\ \hline
\end{tabular}
}
\vspace{-0.5 cm}
\end{table*}
\\
\emph{Field-Programmable Gate Arrays} (FPGAs), rewarded by a high degree of parallelism to accelerate these applications, have been augmented with hard-wired DSP blocks to excel multiplications. Nevertheless, in spite of their advantages, hosting off-the-shelf fixed-precision DSP blocks falls short on fulfilling design requirements in a variety of domains. Beside being unable to perform division, some shortcomings that testify on their inefficiency are: 1) their fixed locations in FPGAs impose routing complexity and often results in degraded performance of some circuits \cite{kuon2007measuring} (and Viterbi decoder, Reed-Solomon and JPEG encoders discussed in \cite{ullah2018area}); 2) unable to be efficiently-utilized for multiplication precision below 18-bit \cite{boutros2018embracing, lee2018double}
(the comparable performance and better energy-efficiency of small-scale LUT-based multipliers over DSP blocks further encourages their deployment in
e.g. neural networks) 
3) their limited ratio versus LUTs ($<$0.001) in multiplication-intensive applications or concurrently executing programs. This forces designers to utilize 
soft- \emph{Intellectual Property} (IP) versions of multipliers and dividers provided by major FPGA vendors such as Xilinx and Intel \cite{XilinxMultIpCore, XilinxDivIpCore}. 

Most of the architectures in the literature,
approximate fixed word-length multipliers or dividers, or SIMD multipliers \cite{leapASPDAC, saadat2018minimally, saadat2019approximate, osorio2019truncated}.
These techniques are not generic since approximation principles (as defined for ASIC) neglect differences in the underlying reconfigurable infrastructure and yield insignificant improvements when directly synthesized and ported to FPGAs \cite{ullah2018smapproxlib}. Few designs have targeted FPGAs which are either \emph{approximate SISD} \cite {ullah2018area, ullah2018smapproxlib} or \emph{accurate SIMD} multipliers \cite{perri2006simd, perri2004variable, lanuzza2005low, khan2009reconfigurable, purohit2008power}.
Moreover, lack of support for division in such architecture imposes substantial overhead on the
design.
This highlights the need for exploring novel avenues to provide a roadmap enabling approximate SIMD-fashion multipliers/dividers, specifically for FPGAs. 

This paper presents for the first time an \emph{\underline{SI}MD approximate soft \underline{M}ultiplier-\underline{Div}ider targeted for FPGAs with Tunabl\underline{e} accuracy}\,--\,\pname. The proposed hybrid architecture not only eliminates the need of reconfiguration,
but can also support both multiplication and division, for the first time, in an integrated hardware with the word-lengths of 8-, 16-, and 32-bits.
We build our architecture based upon Mitchell's algorithm which translates multiplication into addition. This translation results in the simplest linearly-approximated logarithmic multiplier \cite{mitchell1962computer}. In addition, by altering only the additions to subtractions, division can be derived with the same steps.
These translations enable resource-saving, and perfectly fit FPGAs as they are already equipped with fast carry chains hardened to accelerate addition and subtraction. 
\vspace{-0.1cm}

\begin{itemize}[leftmargin=*]
\item \textbf {First integrated approximate multiplier-divider}. We first propose a novel multiplier and a divider
(specially divider with 4.6$\times$ less energy and \texttildelow~\!\!4$\times$ higher speed than accurate version). Afterwards, 
we architect a hybrid design that can be used in either functionality without the need of reconfiguration with still less energy and delay than accurate multiplier.
\item \textbf{An SIMD architecture for proposed multiplier-divider, customized for FPGAs}. To speedup original Mitchell's algorithm and adapt SIMD approach, we propose a 4-bit \emph{Leading-One Detector} (LOD),
which uses two 6-LUTs instead of large priority encoders, used in previous studies. 
Moreover, by adding one controlling signal, our proposed design also successfully implements an approximate SIMD divider, altogether smaller than accurate multiplier.
\item \textbf{Tunable accuracy using novel light-weight error- reduction scheme}.
We use solely one 6-LUT for determining each bit of 64 error-reduction terms and minimal extra circuitry for their addition to
Mitchell's multiplier/divider of arbitrary size.
Our error-coefficients are added with the same LUTs and their associated fast carry chains,
already used for the addition step of Mitchell's algorithm.
This addresses the prolonged critical path in cutting-edge 
approaches \cite{saadat2018minimally, saadat2019approximate, low2015unified}.
We are also able to increase accuracy by one-bit and limit error to a desirable bound using one more LUT ($99.2\%$ accuracy with eight LUTs).
\end{itemize}
\vspace{-0.2cm}
\vspace{-0cm}
\section{Related Work} \label{sec:related}

\textbf{SISD Approximate Multiplier/Divider}:
Studies in this category
employ:
1) LSBs truncation (imposes $>$4\% error in divider).
compared to multiplier. 2) Hierarchical integration of inexact multipliers
(error can drastically accumulate when
deployed in MSBs). 3) Use approximate add/sub for mul/div which offers limited resource saving. 4) Multiplying rounded dividend with reciprocal of divisor.
5) Exploit approximate mul/div algorithms based on LOD.
Such ASIC-evaluated approaches have provided smaller gains in FPGAs. Ultimately, penalty of separate resources for
multiplier and divider still exists.
Shortcomings
of Mitchell-based designs: 1) approximating log of inputs individually,
neglects magnitude of error after multiplication.
2) Lengthened critical path, as selection of error-coefficient depends on the intermediate result of Mitchell's algorithm. 3) Many overflow cases after adding error-reduction term.

\textbf{SIMD Accurate/Approximate Multiplier}:
Authors in \cite{lee2018double, boutros2018embracing} 
have shown performance/energy improvements in FPGA-based DNNs by modifying ASIC-based DSP block to perform double approximate multiplications with a common operand.
Recently, \cite{osorio2019truncated} has proposed an approximate SIMD design (using 8x8 truncated multipliers) for ASIC platforms.
Targeting FPGAs, few works have presented SIMD soft multipliers \cite {perri2004variable, perri2006simd, purohit2008power, lanuzza2005low} that implement accurate 8- to 32-bit multiplication.\\
our work distinguishes itself from SoA as we introduce the first hybrid multiplier/divider with tunable accuracy. Both SISD and SIMD versions are smaller than an accurate multiplier, without the need for reconfiguration or changing architecture of FPGAs
SIMDive can speed up execution of many applications featuring data parallelism while also
reducing the share of energy expenditure by coalescing multiple memory accesses,
both of which increase computational efficiency.



\section{\vspace{0.05cm} Proposed Architecture}  \label{sec:proposed} 
\subsection{\!Preliminaries: \,\!Mitchell’s \,\!Algorithms}
In the binary representation of $N$-bit integer $A$ (Eq. \ref{equation1}), $k$ reveals the position of leading one. The rest of the bits (from position $k-1$ to $0$) are considered as the fractional part, i.e. x, $0\leq x<1$.

\setlength{\abovedisplayskip}{-5pt}
\vspace{-0.3cm}

\footnotesize
\begin{equation}
 \begin{multlined}
\label{equation1}
  A=2^k\!+\sum_{i=0}^{k-1}2^ib_i\!=\!2^k(1+x)\xRightarrow{e.g.} \!\!43=2^5(1+0.01011)_2,10=2^3(1+0.01)_2
     \end{multlined}
\end{equation}
\normalsize
\vspace{-0.3cm}

In linear mathematics, $log_2(1 + x)$ is approximated to $x$ for this range; therefore, the approximate log value of input $A$ is: 
\vspace{-0.3cm}

\begin{equation}
\label{equation2}
\footnotesize
\!Log_2(A) \!\simeq\! k+x 
\footnotesize
\Rightarrow \!\!Log_2(43)\simeq(101.01011)_2, Log_2(10)\simeq(11.01)_2\!
\end{equation}
\vspace{-0.4cm}

In the same manner for second input, summation (subtraction) of two parts is obtained in Eq. \ref{equation3_1} (Eq. \ref{equation3_2}).
\vspace{-0.1cm}

\footnotesize
{
\begin{equation}
\label{equation3_1}
\widetilde{Log_2}(\tilde{P})\!=\!(k_1\! + \!k_2)\!\!\, + \!\!\,(x_1\!+\!x_2)
\!\Rightarrow\!\! \,K_{s}\!=\!(1000)_2,\, X_{s}\!=\!(0.10011)_2\, \end{equation}

\vspace{-0.2cm}

\begin{equation}
\label{equation3_2}
\!\widetilde{Log_2}(\tilde{D})\!\,=\,\!(k_1\!-\!k_2)\!\!\,+ \!\!\,(x_1\!-\!x_2)
\!\Rightarrow\, \!\! K_{s}\!=\!(10)_2,\!\, X_{s}\!=\!(0.00011)_2\!\!
\end{equation}

}
\normalsize{
Finally, by applying anti-log, binary representation of approximate product (quotient) are derived by Eq. \ref{equation4_1} (Eq. \ref{equation4_2}): 
}

\vspace{-0.2cm}

\footnotesize
{
\begin{equation}
\label{equation4_1}
\begin{multlined}
\tilde{P}=\left\{\begin{matrix}
 2^{k_1+k_2}(1+x_1+x_2), & x_1+x_2<1 \\  
 2^{k_1+k_2+1} (x_1+x_2),  & x_1+x_2\geq1
\end{matrix}\right.
\\
\Rightarrow \;\; \tilde{P} = (110011000)_2 = 408,  \;\;  P_{accurate}= 430
 \end{multlined}
\end{equation}

\vspace{-0cm}

\begin{equation}
\label{equation4_2}
\begin{multlined}
\tilde{D}=\left\{\begin{matrix}
 2^{k_1-k_2-1}(2+x_1-x_2), & x_1-x_2<0 \\  
 2^{k_1-k_2} (1+x_1-x_2),  & x_1-x_2\geq0
\end{matrix}\right.
\\
\Rightarrow \,\, \tilde{D} = (100)_2 = 4,  \;\;  D_{accurate}= 4    \;\;\;\;\;\;\;\;\;\;
 \end{multlined}
\end{equation}
}

\vspace{-0.3cm}

\normalsize

\begin{figure*}[t]
 \centering
  \includegraphics[width=\textwidth]{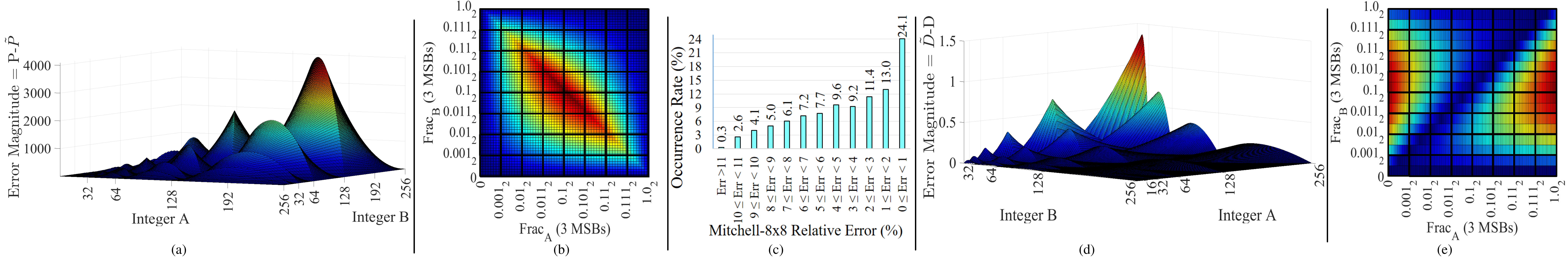}
     \vspace{-0.6cm}
 \caption{Mitchell's error: (a) integer multiplier, (b): top view of multiplier relative error in each power of two interval (same for any size of multiplier) (c): relative error distribution in multiplier, (d): integer divider, (e): top view of divider relative error in each power of two interval (same for any size of divider) }\label{fig:3d_view_error_distribution}
 \vspace{-0.3cm}
\end{figure*}

\begin{figure}[t]
 \centering
  \includegraphics[width=0.48\textwidth]{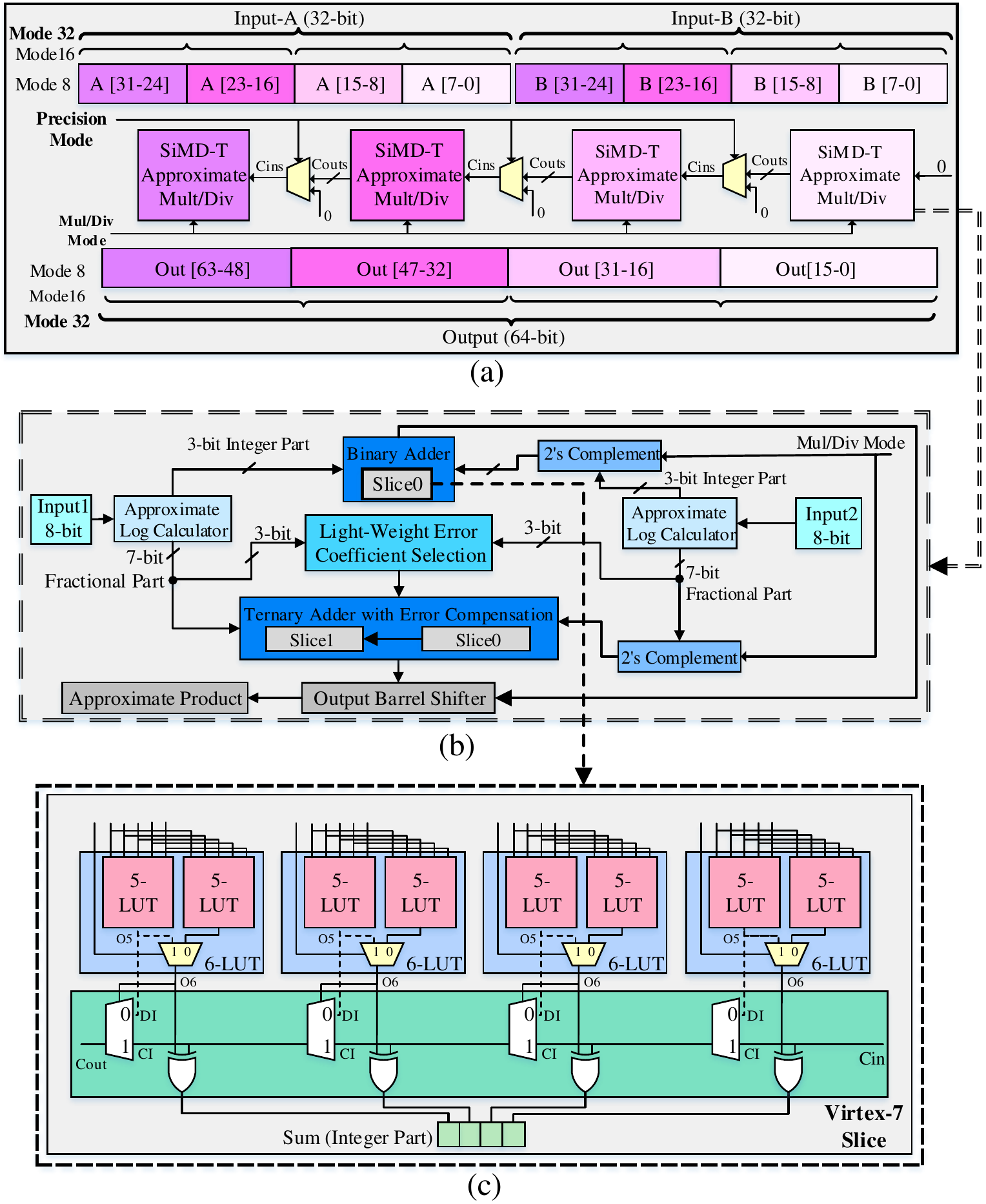}
     \vspace{-.7cm}
 \caption{(a) \pname~Structure, (b) Proposed 8-bit multiplier/divider,\\(c) Virtex-7 slice (used for addition/subtraction of integer/fractional parts)}\label{fig:proposed_multiplier}
  \vspace{-0.7cm}
\end{figure} 
\vspace{-0.3cm}

\subsection {\!\!\!\mysize \pname: Approximate SIMD Multiplier-Divider}
The overall structure of proposed SIMD multiplier-divider is illustrated in Fig. \ref{fig:proposed_multiplier}. Controlling signals \emph{precision} and \emph{Mul/Div mode} shown in Fig. \ref{fig:proposed_multiplier} (a),
serve to establish diverse sub-word size and functionalities of each module, respectively. We used the default one-hot encoding preferred by FPGA manufacturers as it is proven to be more resource-efficient than binary encoding for reconfigurable fabrics \cite{ashenden2010designer}.
One-hot encoding scheme also enables:
1) supporting \emph{mixed-precision} and \emph{mixed-functionality}
for the first time: the proposed SIMD multiplier-divider can either operate as a single 32x32 unit or be decomposed into a twin 16x16, one 16x16 and two 8x8, or quad 8x8 units each of which can act separately as multiplier or divider.
Supporting dynamic mixed-functionality in our design eliminates the need of separate resources/the overhead of reconfiguration and accommodates both operations in a single module. This feature which makes it stand out from previous works is of great interest, especially in multiplication-intensive workloads with fewer division (appealing for both DNN and multimedia). Moreover, in case of sub-word parallel processing, a complete 32-bit \pname~unit is not occupied for a division and it can also orchestrate multiplication if needed.
2) The separate \emph{data-size} signals can also be used to power-gate each sub-unit.

Referring to Fig. \ref{fig:proposed_multiplier} (b), detection of leading one is orchestrated in parallel in our proposed approximate log calculator for each 4-bit segment of the input by using two 6-LUTs: the first LUT acts as a zero-detection flag and detects whether the four bits are zero.
The second 6-LUT (used as two 5-LUTs) is directly configured in such a way to reveal the position of leading-one (0 to 3). Afterwards, depending on the required operand word-length, 
integer and fractional parts are determined based on the most significant non-zero segment.
Based on our analysis, 4-bit segmentation renders the best energy-delay product for 32-bit SIMD architecture.
Afterwards, each 4-bit addition for integer and fractional parts shown in Fig. \ref{fig:proposed_multiplier} (b) are fulfilled by a Virtex-7 slice. As shown in part (c) of this figure, each slice includes four 6-LUTs and its associated fast carry chains, together implement a CLA. Extending the 8-bit addition to 16- and 32-bits in our SIMD architecture is also easily achieved by connecting the $C_{out}$ from previous adder to the $C_{in}$ of next adder (handled by yellow multiplexers in part (a) of this figure).
Division is also performed by altering additions to subtractions (Eq. \ref{equation4_1} and Eq. \ref{equation4_2}). We support this operation by a 2's complement module which calculates the negative of the second input before feeding it to the adder.  

\subsection{\mysize {\!\!\!Proposed Light-Weight Error-Reduction Scheme}}
Mitchell's error for multiplier and divider (8-bit) is plotted by a heat map provided in Fig.\ref{fig:3d_view_error_distribution},
through which four points can be observed: 

\vspace{-0.1cm}
\footnotesize
\begin{equation}
\label{equation5_1}
E_P=P-\tilde{P}=\left\{\begin{matrix}
 2^{k_1+k_2}(x_1x_2), & x_1+x_2<1\\ 
 2^{k_1+k_2} (1-x_1-x_2+x_1x_2),  & x_1+x_2\geq1
\end{matrix}\right.
\end{equation}
\normalsize

\vspace{0.1cm}
\footnotesize
\begin{equation}
\label{equation5_2}
E_D=D-\tilde{D}=\left\{\begin{matrix}
 2^{k_1-k_2}\, \frac{(x_1(x_2-1)+x_2-(x_2)^2)}{2(1+x_2)}, & x_1-x_2<0\\ 
 2^{k_1-k_2} \, \frac{(x_1x_2-(x_2)^2)}{1+x_2},  & x_1-x_2\geq0
\end{matrix}\right.
\end{equation}

\normalsize

\begin{itemize}[leftmargin=*]
\item Different error magnitude in each power-of-two interval (Fig.\ref{fig:3d_view_error_distribution} (a), (d)) demonstrates that adding a single correction term to the output cannot fit for all multiplier/divider sizes.
\item Eq.\! \ref{equation5_1} and Eq. \ref{equation5_2} prove proportional replication of error in each power-of-two:
irrespective of $k_1$ and $k_2$, unique
schemes for each of multiplier/divider may fit all sizes and they can be added to fractional part before scaling to save more resources.
\item Fig.\ref{fig:3d_view_error_distribution} (b), (e) exhibit the non-uniform, but symmetrical error distribution: errors tends to be the same at the beginning and end of each power-of-two interval, encouraging the same reduction approach for all multiplier or divider sizes. 
\item Finally, Fig.\ref{fig:3d_view_error_distribution} (c) shows diverse distribution of relative error. This means employing a single error-coefficient to the whole interval (as proposed in SoA MBM \cite{saadat2018minimally} and INZeD \cite{saadat2019approximate}), is not efficient and results in many output overflow cases. 

\end{itemize}



\vspace{-0.1cm}
Coalescing the insights from above points incentivize using multiple error-reduction terms appropriately opted based on summation of fractional parts to cope with overflow and yet enduring minimal latency overhead. In our analysis, we attempted to optimize two factors:
1) \emph{error\,\,magnitude\,$\times$\,error\,\,distribution} in each region (can be estimated as the integral of error-magnitude of the region).
2) partitioning overhead which depends on the number of MSBs checked in fractional parts (e.g., checking up to 11 bits as proposed in \cite{low2015unified} would thwart the gain of approximation).
Elaborately building our architecture upon these observations and benefiting from the underlying FPGA structure, we have proposed a novel error-reduction scheme based on efficiently utilizing 6-LUTs: we assign 3 MSBs of each fractional part to LUT inputs which is responsible for calculating one bit of the error-coefficient. As illustrated in Fig. \ref{fig:3d_view_error_distribution} (b) and (e), the squarish region for all combination of inputs is subdivided to 64 sub-regions by merely using 3 MSBs of inputs. We assign a distinctive coefficient to each of these regions, representing their average error for each sub-interval.
64 output entries of $i^{th}$ LUT determine $i^{th}$ bit of these 64 coefficients in binary representation. Therefore, using only one LUT, we can efficiently determine one bit of 64 error-coefficient terms.
Having 64 coefficients appropriately calculated based on the combination of both operands addresses both drawbacks discussed in Section \ref{sec:related} (neglecting magnitude of error due to separately approximating each operand and overflow cases).
Such many overflow cases cases neither are measured, nor handled in\cite{saadat2018minimally, saadat2019approximate}. In our scheme, they are alleviated by assigning an appropriate coefficient to each pair of input.

LUTs and their associated fast carry chain in Xilinx UNISIM library \cite{Xilinx7SeriesGuide} can be configured to implement a ternary adder. This perfectly suits our error-healing approach as \emph{we are able to combine the process of adding error-reduction term with fractional parts within the same resources in a single step}. Regardless of adder size, only one more bit at MSB is needed in ternary addition (compared to binary version), since frac1$_i$+frac2$_i$+error\_term$_i$+$C_{in}$ ($C_{out}$ of from previous bit) may result in 3 bits, necessitates a more LUT at the end of the chain.
Moreover, the delay of FPGA primitives is fixed and adding error-reduction term at the same time when fractional parts are added keeps the overall delay of the circuit nearly untouched.

\vspace{-0.2cm}
\subsection{Applicability to other FPGAs}
Our approach is easily applicable to various FPGAs without the need for architectural modification. It can achieve even better accuracy in higher-bit LUTs (e.g., 8-bit ALMs in Intel's Stratix and Arria series): considering 4-bits of each fractional parts will provide 256 sub-regions. Therefore, solely one LUT can enable 256 error-reduction coefficients which can significantly improve accuracy\footnote{Ternary addition also can be implemented efficiently in Stratix and Arria series with higher frequency \cite{IntelALM}}. Our error-analysis reveals that the proposed light-weight error-reduction approach in this paper \textit{can significantly reduce average relative error to\;$<$\,0.1\%} using 8-LUTs).


\section{Results and Discussion
\label{sec:results}}

\subsection{Experimental Setup}
We have evaluated \pname~against five SIMD and SISD accurate and approximate cutting-edge multipliers and dividers:
performance-optimized accurate IPs of multiplier
\cite{XilinxMultIpCore} and divider
\cite{XilinxDivIpCore},
provided by Xilinx Vivado, Mitchell \cite{mitchell1962computer}, SoAs MBM  \cite{saadat2018minimally}, INZeD \cite{saadat2019approximate} and AAXD dividers \cite{saadat2019approximate} as they have the best resource-error trade-off when compared to the rest of designs (\cite {jiang2018adaptive, 8464807, vahdat2017truncapp, hashemi2016low}), 
CA \cite{ullah2018area} (based on approximate 4x4 multipliers) customized for FPGAs, and truncated multiplier (with 7x7 or 15x7 as the basic multiplier, the more accurate one is also exploited in SIMD structure). 
Note, 
hierarchical SIMD divider is not mathematically feasible by
decomposing large one to small instances.
Even implementing division with reciprocal-function would be approximate and divider IP is still needed for its implementation. 
However, by exploiting Mitchell's algorithm in our proposed \pname, we have made its approximated SIMD mode possible (as division is translated to subtracting two pairs of numbers and then a simple shifting).

All designs are implemented from scratch in 16-bit SISD mode to provide insight about their individual resource footprints and accuracy. Afterwards, they have been exploited in 32-bit SIMD architecture. 
Each circuit is coded in VHDL and synthesized and implemented by Vivado 17.4 for Virtex-7 VC707 FPGA.
Area, throughput, and power are reported from Vivado and Power Analyzer simulations over $10^6$ for SISD and $10^9$ inputs for SIMD mode uniformly distributed in a random order in the whole 16- and 32-bit interval, respectively.
For precise estimations,
throughput and energy dissipation are calculated based on the total execution time and power consumed for all the inputs fed to the \pname. We used the cost function defined by \cite{7762134}, i.e., $Area\times Energy\times Delay$/$(1-NED)$,
Where Normalized Error Distance (NED) is the error distance for all inputs divided by maximum error.
Design metrics are also reported separately since: 1) the weighted product of quality-resource metrics may not always be an appropriate figure of merit and lacks distinctiveness \cite{rehman2016architectural}.
2) Depending on designer/application preference, each metric can have more importance over the other. 
The behavioral models of multipliers/dividers are also developed in MATLAB, C++, and Python to calculate average absolute relative error and peak absolute relative error (referred to as relative and peak errors, respectively) for all possible multiplier inputs. 
We have also deployed \pname~during the inference phase of a lightweight Artificial Neural Network (ANN), Gaussian Image Smoothing, and Multiply-based Image blending applications to test the applicability of \pname~in real-world applications.
\begin{table}[]
\huge
\centering
\caption{{Design metrics in SISD multipliers (16x16) and dividers (16/8)}}
\vspace{-0.3cm}
\label{table:result_16}
\begin{threeparttable}
\resizebox{0.5\textwidth}{!}{

\begin{tabular}{c|cccccccc}

\toprule
 & \multicolumn{1}{c}{\textbf{\HUGER SISD Circuit}}                                                                                                                                                              & \multicolumn{1}{c}{\textbf{\begin{tabular}[c]{@{}c@{}}\HUGER Area\\ \HUGER (6-LUT)\end{tabular}}} & \multicolumn{1}{c}{\textbf{
 \begin{tabular}[c]{@{}c@{}}\HUGER Delay\\ \HUGER (nS)\end{tabular}}} & \multicolumn{1}{c}{\textbf{\begin{tabular}[c]{@{}c@{}}\HUGER Power\\\HUGER (mW)\end{tabular}}} & \multicolumn{1}{c}{\textbf{\begin{tabular}[c]{@{}c@{}}\HUGER Energy\\ \HUGER (uJ)\end{tabular}}} & \multicolumn{1}{c}{\textbf{\begin{tabular}[c]{@{}c@{}}\HUGER ARE$^1$  \\\HUGER (\%)\end{tabular}}} & \multicolumn{1}{c}{\textbf{\begin{tabular}[c]{@{}c@{}}\HUGER PRE$^2$ \\\HUGER (\%)\end{tabular}}} & \multicolumn{1}{c}{\textbf{\begin{tabular}[c]{@{}c@{}}\HUGER CF$^3$ \HUGER \end{tabular}}} 

\\ \midrule


\multirow{7}{*}{\textbf{MUL}}
& \multicolumn{1}{c}{ \HUGER Accurate IP~\cite{XilinxMultIpCore}}                                                                               \HUGER                                                                     &  \multicolumn{1}{c}{\HUGER 287}                                                       &  \multicolumn{1}{c}{\HUGER 6.4}                                                         &  \multicolumn{1}{c}{\HUGER 47.8}                                                    & \multicolumn{1}{c}{\HUGER 306}                                                     & \multicolumn{1}{c}{\HUGER -}                                                                   & \multicolumn{1}{c}{\HUGER -}& \multicolumn{1}{c}{\HUGER 1}                 \\

& \multicolumn{1}{c}{ \HUGER CA {\cite{ullah2018area}}
}            &  \multicolumn{1}{c}{\HUGER 245}                                                      &  \multicolumn{1}{c}{\HUGER 6.8}                                                         &  \multicolumn{1}{c}{\HUGER 46.6}                                                    &  \multicolumn{1}{c}{\HUGER 317}                                                     &  \multicolumn{1}{c}{\HUGER 0.3}                                                                    &  \multicolumn{1}{c}{\HUGER 19.04}        &  \multicolumn{1}{c}{\HUGER 0.68}        
\\ 
& \multicolumn{1}{c}{ \HUGER Trunc (four 7x7) 
 }                                                                                                                 &  \multicolumn{1}{c}{\HUGER 200}                                                       &  \multicolumn{1}{c}{\HUGER 5.6}                                                         &  \multicolumn{1}{c}{\HUGER 40.5}                                                    & \multicolumn{1}{c}{\HUGER 227}                                                     & \multicolumn{1}{c}{\HUGER 2.35}                                                                    &  \multicolumn{1}{c}{\HUGER 100}   &  \multicolumn{1}{c}{\HUGER 0.46}                                                           
\\ 
& \multicolumn{1}{c}{\HUGER  Trunc (two 15x7) 
 }                                                                                                                 &  \multicolumn{1}{c}{\HUGER 202}                                                       &  \multicolumn{1}{c}{\HUGER 5.8}                                                         &  \multicolumn{1}{c}{\HUGER 42}                                                    & \multicolumn{1}{c}{\HUGER 245}                                                     & \multicolumn{1}{c}{\HUGER 1.19}                                                                    &  \multicolumn{1}{c}{\HUGER 100} &  \multicolumn{1}{c}{\HUGER 0.51}                                                           
\\ 
 & \multicolumn{1}{c}{\HUGER  Mitchell \cite{mitchell1962computer}}                                                                                                                                              & \multicolumn{1}{c}{\HUGER 174}                                                       &  \multicolumn{1}{c}{\HUGER 4.7}                                                         &  \multicolumn{1}{c}{\HUGER 35.5}                                                    &  \multicolumn{1}{c}{\HUGER 167}                                                     & \multicolumn{1}{c}{\HUGER 3.85}                                                                    &  \multicolumn{1}{c}{\HUGER 11.11}                                                              &  \multicolumn{1}{c}{\HUGER 0.43}  
\\ 
& \multicolumn{1}{c}{\HUGER  MBM \cite{saadat2018minimally}}                                                                                                                                                           &  \multicolumn{1}{c}{\HUGER 186}                                                       &  \multicolumn{1}{c}{\HUGER 5.4}                                                         &  \multicolumn{1}{c}{\HUGER 36.3}                                                    &  \multicolumn{1}{c}{\HUGER 196}                                                     &  \multicolumn{1}{c}{\HUGER 2.63}                                                                    & \multicolumn{1}{c}{\HUGER 8.81}                                                      & \multicolumn{1}{c}{\HUGER 0.41}                                                             
\\ 
& \multicolumn{1}{c}{\cellcolor{lightgray} \begin{tabular}[c]{@{}c@{}}  {\cellcolor{lightgray} \HUGER Proposed} \end{tabular}}                & \multicolumn{1}{c}{ \cellcolor{lightgray} \HUGER 211}                                                       & \multicolumn{1}{c}{ \cellcolor{lightgray} \HUGER 4.8}                                                         & \multicolumn{1}{c}{\cellcolor{lightgray} \HUGER  37.7}                                                    & \multicolumn{1}{c}{\cellcolor{lightgray} \HUGER 178}                                                     & \multicolumn{1}{c}{\cellcolor{lightgray} \HUGER  0.82}                                                                    & \multicolumn{1}{c}{\cellcolor{lightgray}  \HUGER 4.9} & \multicolumn{1}{c}{ \cellcolor{lightgray} \HUGER 0.34}                                                             
\\ 

\midrule 
\multirow{6}{*} {\textbf{DIV}}

 & \multicolumn{1}{c}{\HUGER  Accurate IP \cite{XilinxDivIpCore}}                                                                                                                                                  &  \multicolumn{1}{c}{\HUGER 168}                                                       &  \multicolumn{1}{c}{\HUGER 21.4}                                                         &  \multicolumn{1}{c}{\HUGER 24.6}                                                    & \multicolumn{1}{c}{\HUGER 526}                                                     & \multicolumn{1}{c}{\HUGER -}                                                                   & \multicolumn{1}{c}{\HUGER -}    & \multicolumn{1}{c}{\HUGER 1}                    
\\

& \multicolumn{1}{c}{ \HUGER AAXD (12/6) \cite{jiang2018adaptive}}                                                                                                                                              & \multicolumn{1}{c}{\HUGER 216}                                                       &  \multicolumn{1}{c}{\HUGER 18.5}                                                         &  \multicolumn{1}{c}{\HUGER 22.8}                                                    &  \multicolumn{1}{c}{\HUGER 412}                                                     & \multicolumn{1}{c}{\HUGER 0.74}                                                                    &  \multicolumn{1}{c}{\HUGER 100}    &  \multicolumn{1}{c}{\HUGER 0.96}                                                              
\\

& \multicolumn{1}{c}{ \HUGER AAXD (8/4) \cite{jiang2018adaptive}}                                                                                                                                              & \multicolumn{1}{c}{\HUGER 148}                                                       &  \multicolumn{1}{c}{\HUGER 11.9}                                                         &  \multicolumn{1}{c}{\HUGER 21}                                                    &  \multicolumn{1}{c}{\HUGER 263}                                                     & \multicolumn{1}{c}{\HUGER 2.99}                                                                    &  \multicolumn{1}{c}{\HUGER 100}   &  \multicolumn{1}{c}{\HUGER 0.3}                                                              
\\

& \multicolumn{1}{c}{ \HUGER Mitchell \cite{mitchell1962computer}}                                                                                                                                              & \multicolumn{1}{c}{\HUGER 119}                                                       &  \multicolumn{1}{c}{\HUGER 5.3}                                                         &  \multicolumn{1}{c}{\HUGER 20.3}                                                    &  \multicolumn{1}{c}{\HUGER 107}                                                     & \multicolumn{1}{c}{\HUGER 4.11}                                                                    &  \multicolumn{1}{c}{\HUGER 13}  &  \multicolumn{1}{c}{\HUGER 0.17}                                                              
\\
 & \multicolumn{1}{c}{\HUGER  INZeD \cite{saadat2019approximate}}                                                                                                                                                           &  \multicolumn{1}{c}{\HUGER 160}                                                       &  \multicolumn{1}{c}{\HUGER 5.9}                                                         &  \multicolumn{1}{c}{\HUGER 22.5}                                                    &  \multicolumn{1}{c}{\HUGER 137}                                                     &  \multicolumn{1}{c}{\HUGER 2.93}                                                                    & \multicolumn{1}{c}{\HUGER 9.5}    & \multicolumn{1}{c}{\HUGER 0.11}                                                             
\\ 
& \multicolumn{1}{c}{\cellcolor{lightgray} \begin{tabular}[c]{@{}c@{}}   \HUGER Proposed\end{tabular}}                & \multicolumn{1}{c}{\cellcolor{lightgray} \HUGER 140}                                                       & \multicolumn{1}{c}{\cellcolor{lightgray} \HUGER 5.4}                                                         & \multicolumn{1}{c}{\cellcolor{lightgray} \HUGER 21.2}                                                    & \multicolumn{1}{c}{\cellcolor{lightgray}\HUGER 114}                                                     & \multicolumn{1}{c}{ \cellcolor{lightgray}\HUGER 0.77}                                                                    & \multicolumn{1}{c}{\cellcolor{lightgray} \HUGER 5.24}  & \multicolumn{1}{c}{\cellcolor{lightgray} \HUGER 0.06}    \\    \midrule   
\multicolumn{2}{c}{ \cellcolor{lightgray} \HUGER  Proposed Integrated Mul-Div} 
& \cellcolor{lightgray}  \HUGER   268       & \cellcolor{lightgray}   \HUGER  5.8   &\cellcolor{lightgray}  \HUGER    44  & \cellcolor{lightgray}  \HUGER  254   & \cellcolor{lightgray} \HUGER 0.82 & \cellcolor{lightgray} \HUGER  5.24 & \cellcolor{lightgray} s\HUGER 0.22   
\\ \bottomrule

\end{tabular}
}

 
\end{threeparttable}
\end{table}

\vspace{-0.3cm}

\begin{table}[]
\large

\centering
\caption{{Design metrics of proposed approximate 32-bit SIMD multiplier-divider and SoA multipliers/dividers implemented in SIMD fashion}}
\vspace{-0.3cm}
\label{table:result_32}
\resizebox{0.5\textwidth}{!}{

\begin{tabular}{c|ccccc}

\toprule
  &          \multicolumn{1}{c}{\textbf{SIMD Basic Block}}    & \textbf{\begin{tabular}[c]{@{}c@{}}Area\\ (LUT)\end{tabular}} & \textbf{\begin{tabular}[c]{@{}c@{}}Throughput\\uS
\end{tabular}} & \textbf{\begin{tabular}[c]{@{}c@{}}Power\\(mW)\end{tabular}}        & \textbf{\begin{tabular}[c]{@{}c@{}}Energy\\(mJ)\end{tabular}}     \\ \midrule
 \multirow{3}{*}{\textbf{MUL}}
 & Accurate Multiplier~\cite{perri2006simd}         &       1125     & 561  &         121          &         862           \vspace{0.05cm}  \\ 
& CA   {\cite{ullah2018area}}         &      1071      &    412      &       106         &       1028       \vspace{0.05cm}    \\ 
& Truncated (using 31x7)             &     970      &    645  &         98       &        608        \vspace{0.05cm} \\
\midrule  \multicolumn{2}{l}{} Accurate Divider  (32-bit, SISD)         &     592      &    22   &    38               &             954        \\ 
 \midrule
 \multirow{3}{*}{\textbf{MUL/DIV}}
 & Mitchell Mul-Div    \cite{mitchell1962computer}    &   782   &    851  &         72           &         339        \vspace{0.05cm} \\ 
& MBM-INZeD    \cite{saadat2018minimally}-\!\cite{saadat2019approximate}     &    910  &  694    &  89.4                   &        515         \vspace{0.05cm} \\ 
 & \cellcolor{lightgray} Proposed \pname &  \cellcolor{lightgray}    834      &    \cellcolor{lightgray} 817    &  \cellcolor{lightgray}      77.5        &      \cellcolor{lightgray}  379           \vspace{0.05cm}   \\ 
\bottomrule
\end{tabular}
}

 \vspace{-0.6cm}
\end{table}

\normalsize

\subsection{Simulation and Synthesis Results}
\!\!Tables \ref{table:result_16} and \ref{table:result_32} summarize design metrics and error analysis. Following conclusions are notable referring to these tables:
\vspace{-0.1cm}
\begin{itemize}[leftmargin=*]
\item \emph{\pname~corroborates its superiority by improving resource consumption}: Comparing 
\pname~with truncated and hierarchical-based counterparts, designed upon incorporating smaller instances, justifies following points: 

\hspace{0.2cm}1)
all resource footprints are improved in Mitchell-based designs compared to the accurate counterparts. In particular, delay and energy are improved by \texttildelow4$\times$ and 4.6$\times$, respectively, in our proposed divider in SISD mode, as compared to accurate counterpart. In contrast, CA \cite{ullah2018area} with hierarchical implementation approach dissipates even more energy with lower throughput than accurate multiplier. 



\hspace{0.2cm}2) Approximation applied on hierarchical multipliers is rewarding in accuracy-resource trade-off only when it is done from scratch for each size.
Otherwise by integrating smaller instances, error can be significantly accumulated as truncated bits are also placed in upper bit-positions. This means that error of CA multiplier would drastically increase
in 32-bit
using smaller instances. 
In contrast, instead of implementing hierarchical approach by connecting approximate modules, we have exploited resource reuse in Mitchell algorithm as much as possible: e.g., 8-bit accurate adders are connected to make a 32-bit instance, or detection of leading one is performed in parallel in each 4-bit segment of inputs. Therefore, no additional error is incurred in our design when used in 16- or 32-bit.

\hspace{0.2cm}3) Through novel light-weight error-reduction scheme specifically customized for FPGAs, \pname~has achieved significant improvement in terms of delay, energy, and accuracy, specially compared to SoA INZeD \cite{saadat2019approximate} (mitchell-based) and AAXD \cite{jiang2018adaptive} (dynamically truncation of division operands).

\hspace{0.2cm}4) Transforming our proposed integrated multiplier-divider from 16-bit SISD to 32-bit SIMD has increased the area by factor of \texttildelow3. This factor is \texttildelow4 for hierarchical designs which are still either SIMD multiplier or SISD divider.
The reason behind this is two-fold: a) as discussed before, our proposed LOD detects position of leading one in each 4-bit segment. This enables resource reuse and imposes small overhead when 32-bit LOD is converted to 8-bit. b) Mitchell's algorithm is also inherently more suited for SIMD architecture, as converting SISD to SIMD in addition steps of the algorithm poses small overhead (modifying 32-bit SISD adder in fractional part to four 8-bit adders in SIMD mode). Overall, when input size is multiplied by a factor of x, resource footprint grows quadratic (x$^2$) in hierarchical multiplication approach, while it increases less aggressively by utilizing our logarithmic designs.

\item \emph{The proposed error-refinement approach surpasses SoAs}: 
By augmenting Mitchell's algorithm with our novel error-reduction schemes independent from the input size, we successfully achieved the lowest peak error among approximate designs (up to 20$\times$). Additionally, average relative error is also limited to $<$\,0.8\%. Although exhaustive 32x32 test is prohibitively time consuming \cite{osorio2019truncated}, average error in Mitchell-based designs will not significantly change in larger bit-width. Moreover, our proposed error-reduction scheme is independent from input-width and as shown by the result, it outperforms MBM design with respect to both error metrics (\texttildelow5$\times$ less average and \texttildelow2.5$\times$ less peak error). Finally, boosting precision in our approach comes with a negligible cost: one more LUT increases the precision of error-coefficient by one bit (as discussed in Section \ref{sec:proposed}.C).

\vspace{-0.1cm}

\end{itemize}

\subsection{ANN and Image Processing Applications}
\begin{figure}[t] 
        \vspace{-0.1cm}
         \subfloat[\scriptsize{Original image}\label{sub_fig1}]{%
       \includegraphics[width=0.116\textwidth]{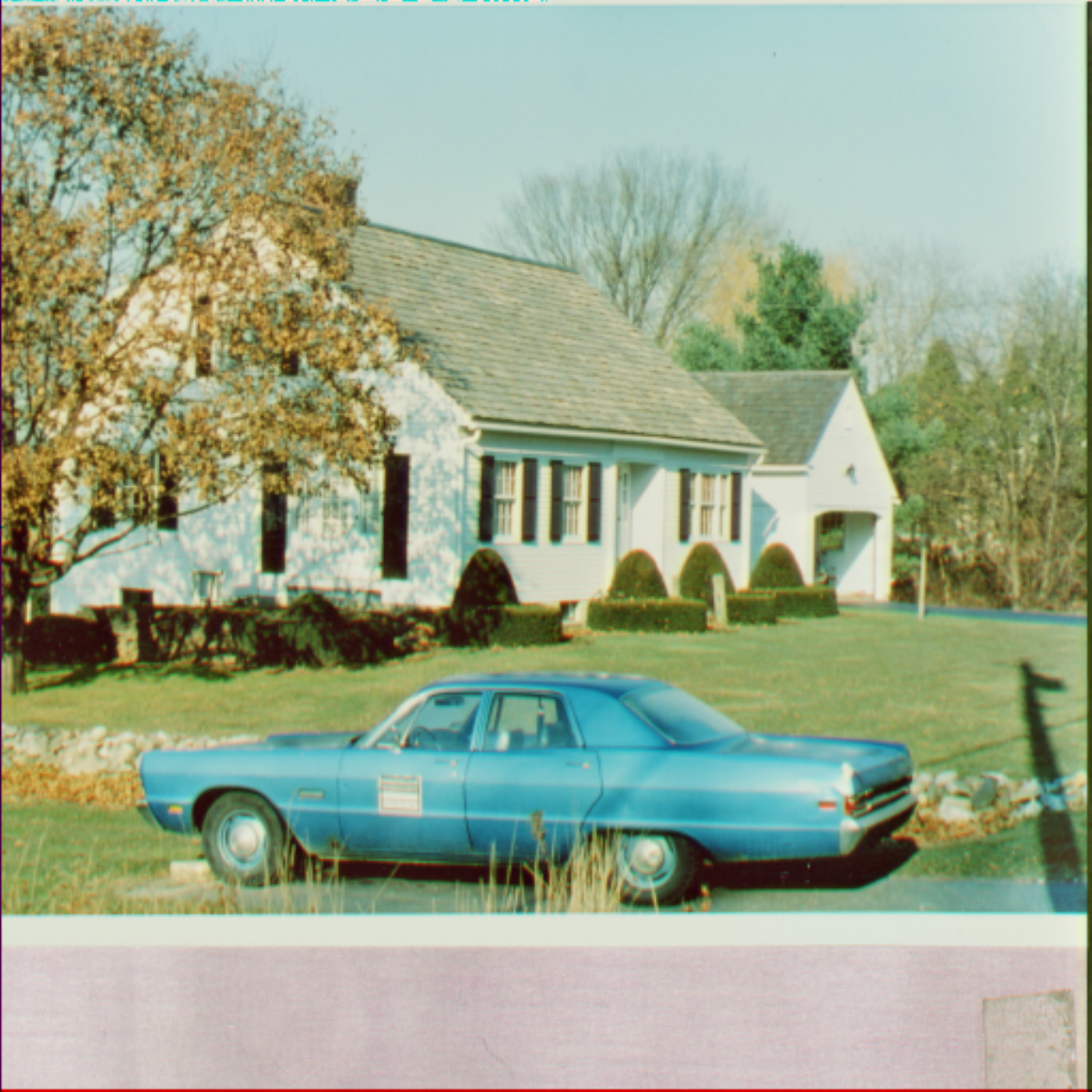}}
    \hfill
  \subfloat[\scriptsize{Accurate}\label{sub_fig2}]{%
        \includegraphics[width=0.116\textwidth]{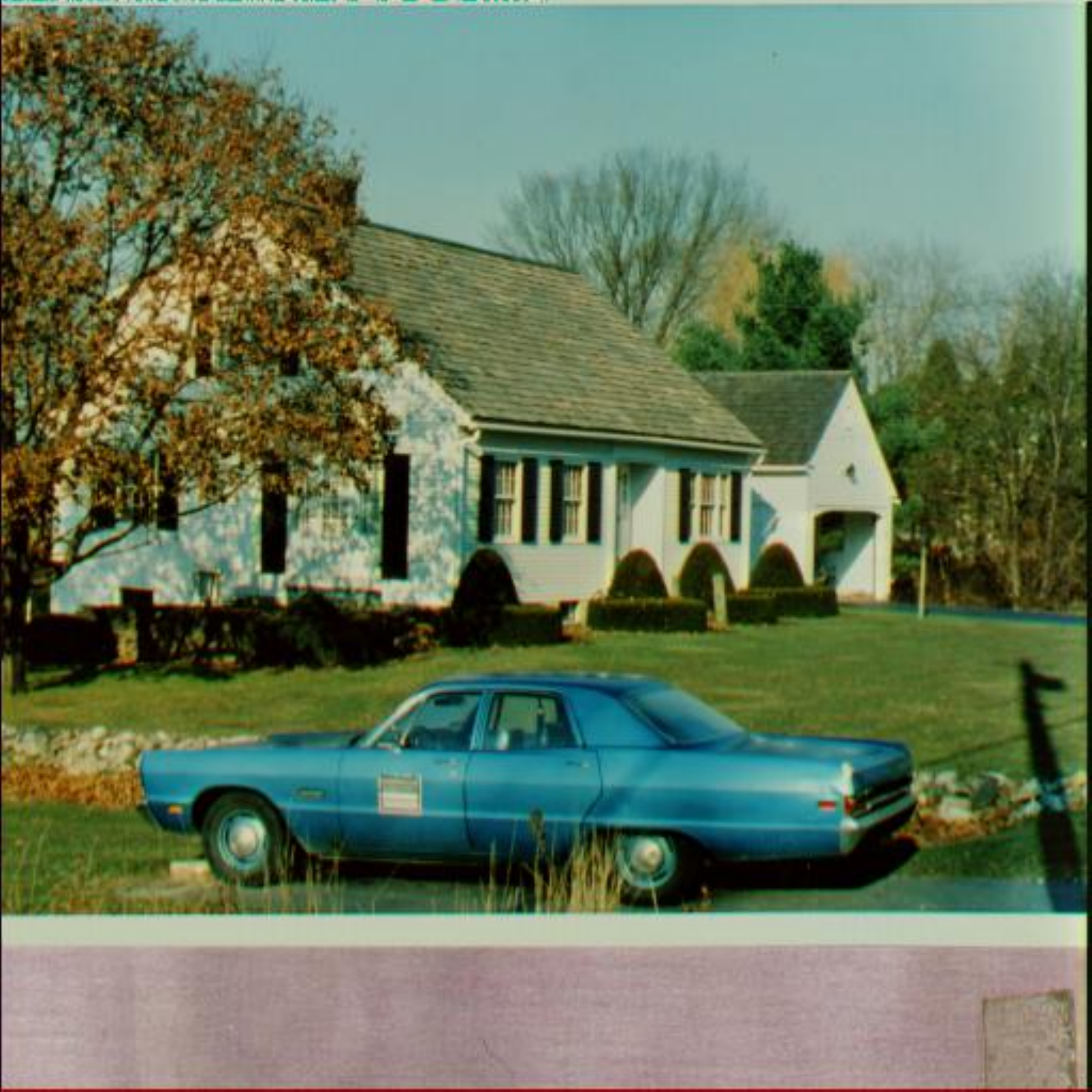}}
        \hfill
      \subfloat[\scriptsize{\!\pname=45.3}\label{sub_fig3}]{%
        \includegraphics[width=0.116\textwidth]{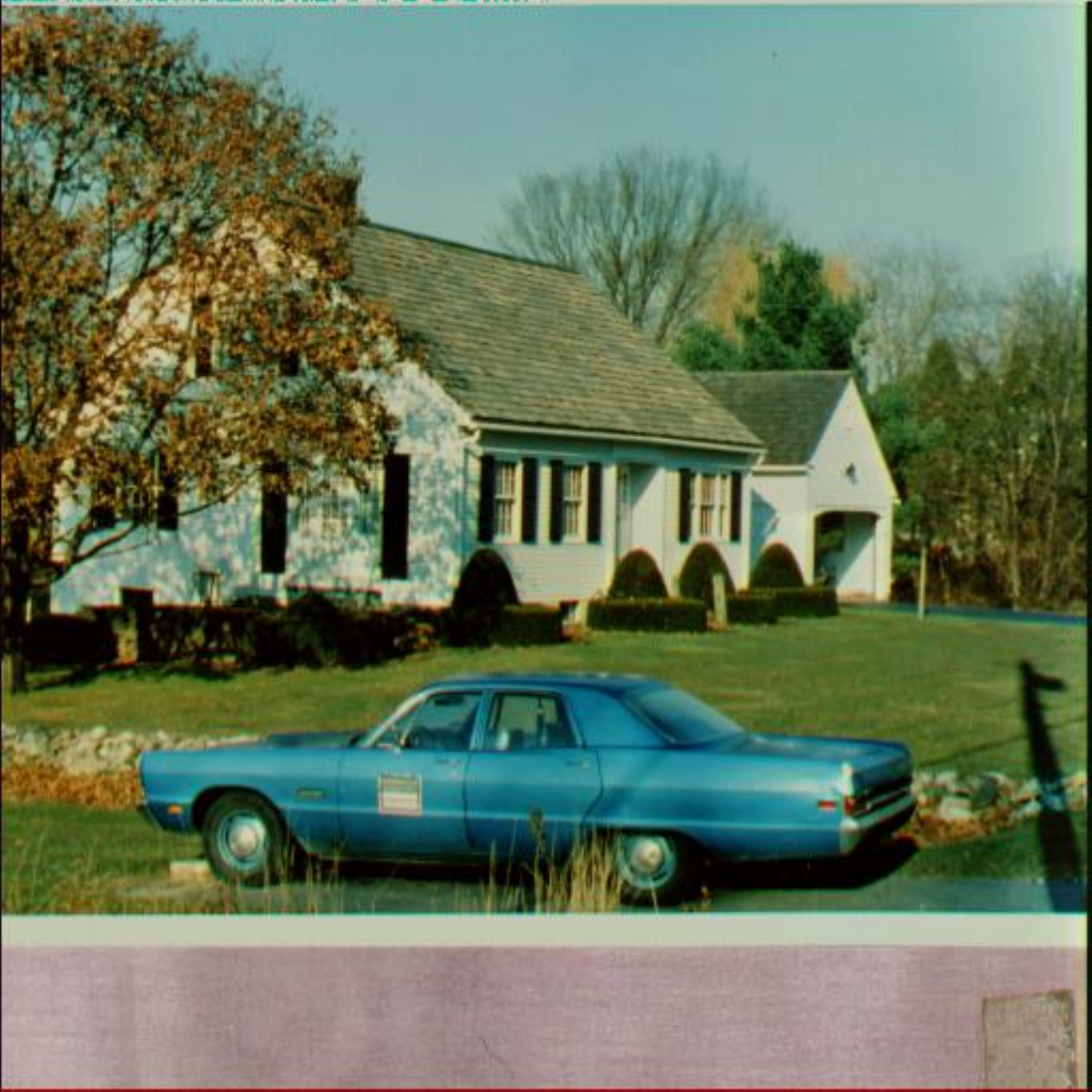}}
            \hfill
      \subfloat[\scriptsize{\!MBM~\cite{saadat2018minimally}=30.7}\label{sub_fig4}]{%
        \includegraphics[width=0.116\textwidth]{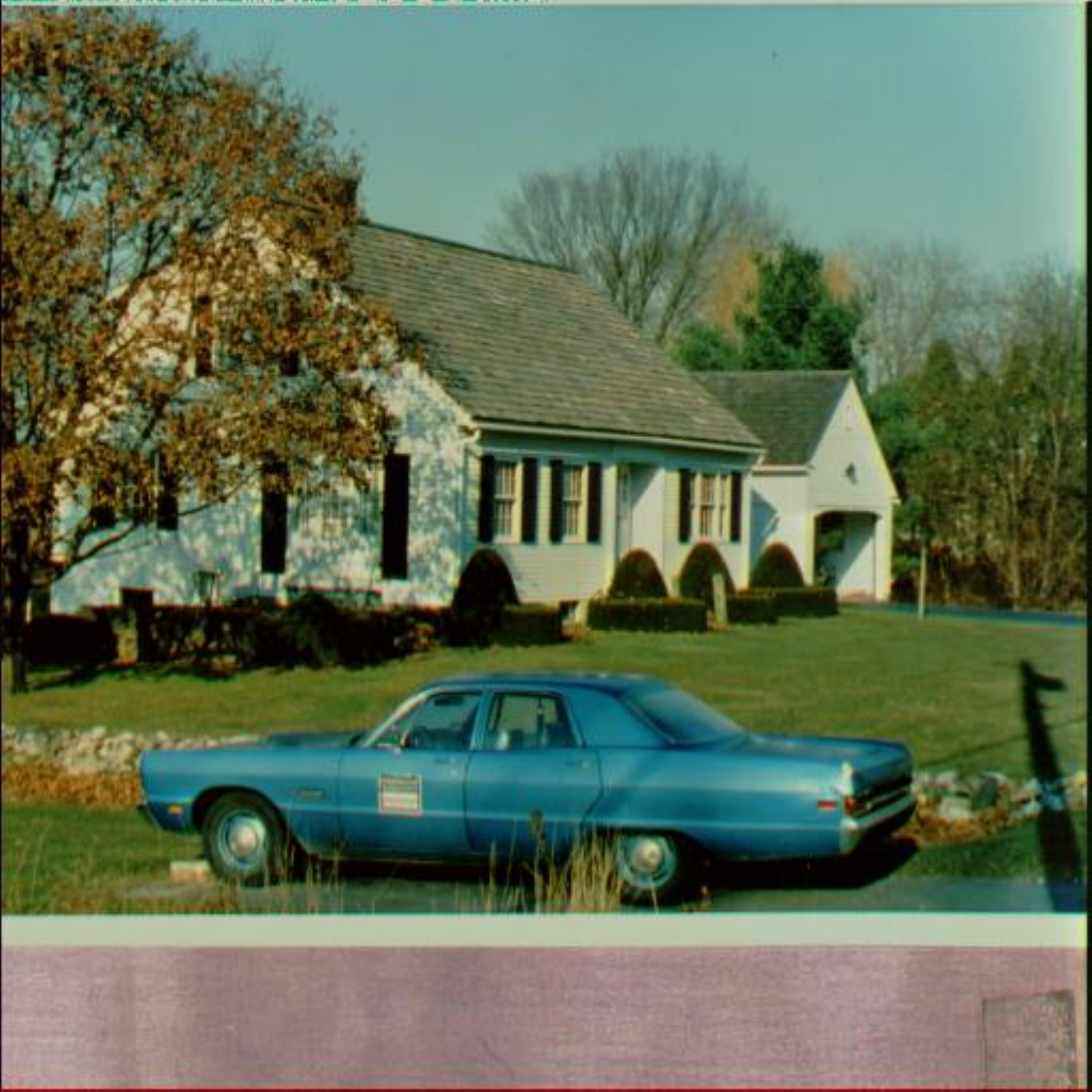}}
       \vspace{-0.3cm}
  \caption{Image blending application with various multipliers (PSNR values are w.r.t. the accurate multiplier-based filter)
 }
  \label{fig:image_blend} 
  \vspace{-0.4cm}
\end{figure}

\begin{figure}[t] 
        \vspace{-0.1cm}
         \subfloat[\scriptsize{\!\,Noise-induced=20.1\!\!\!\!}\label{sub_fig1}]{%
       \includegraphics[width=0.118\textwidth]{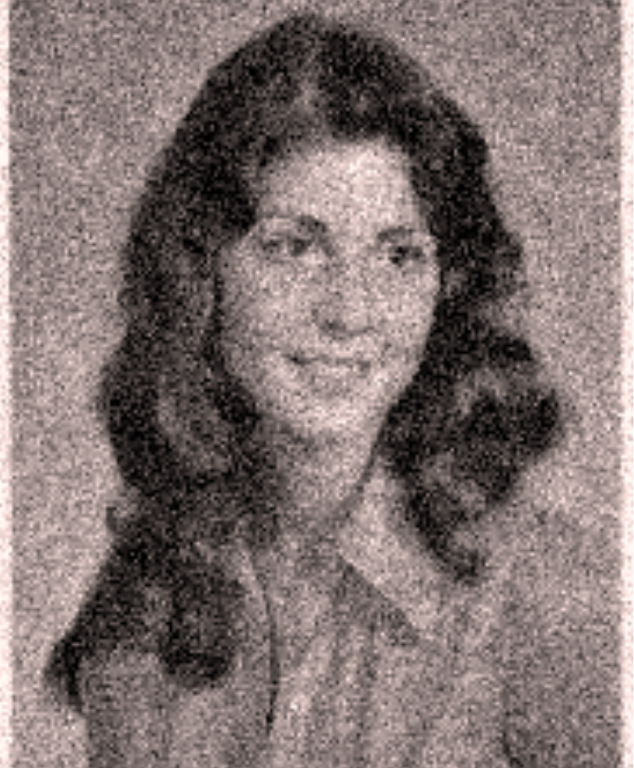}}
    \hfill
  \subfloat[\scriptsize{\!Accurate=24.5\!}\label{sub_fig2}]{%
        \includegraphics[width=0.118\textwidth]{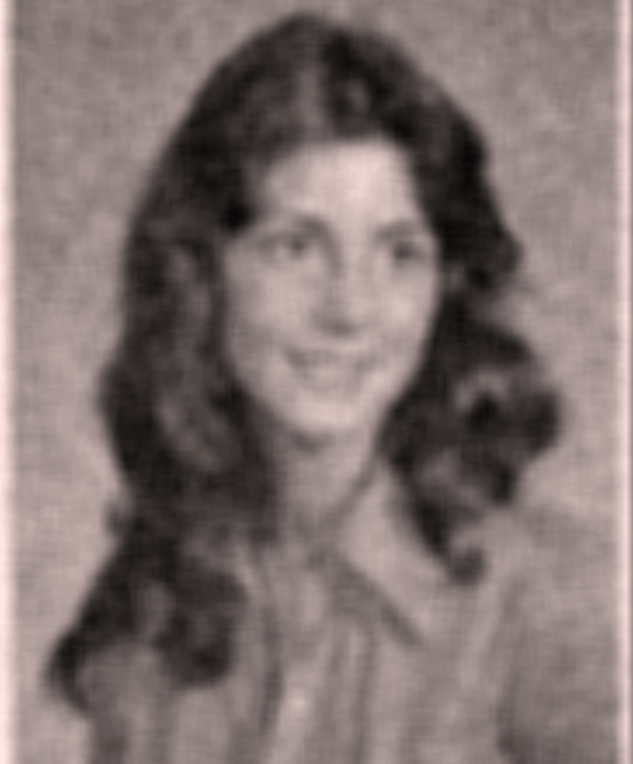}}
        \hfill
      \subfloat[\scriptsize{\!\pname=24, 23.3\!}\label{sub_fig3}]{%
        \includegraphics[width=0.118\textwidth]{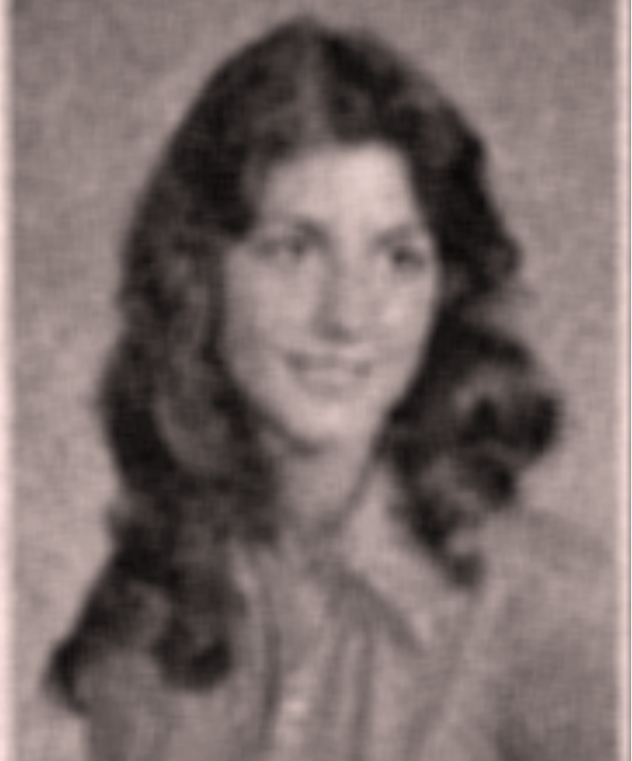}}
            \hfill
      \subfloat[\tiny{\!MBM/INZeD=21.3,\! 20.5}\label{sub_fig4}]{%
        \includegraphics[width=0.118\textwidth]{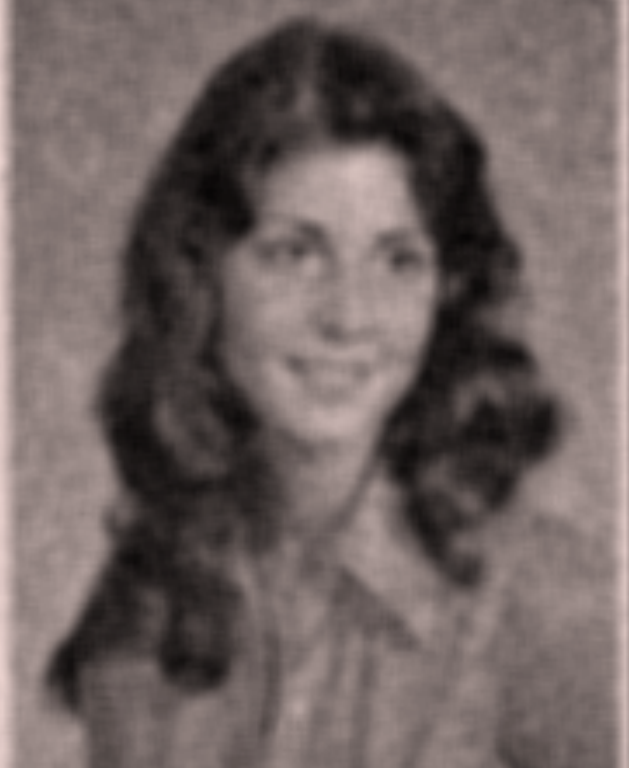}}
       \vspace{-0.3cm}
  \caption{\footnotesize{Gaussian noise removal filter (divider and hybrid mul/div modes, PSNR values are w.r.t the original noise-free image)}
 }
  \label{fig:gaussian} 
  \vspace{-0.5cm}
\end{figure}



For further quality assessment of~\pname~in high-level applications, we considered Image Blending and Gaussian Image Smoothing applications for 
USC-SIPI Database~\cite{uscsipi}.
For Image Blending application, all multipliers in the application have been replaced with approximate versions provided by~\pname~and MBM~\cite{saadat2018minimally}. The average Peak signal-to-noise ratio (PSNR) value produced by \pname-based application
is $46.6$, whereas it is limited to $32.1$ for the MBM counterpart. Fig.~\ref{fig:image_blend} presents visual qualities of two processed images. For Gaussian Image Smoothing filter, we modified the application
for two distinct modes: 1) only approximate version of divider is deployed. 2) Hybrid: both multiplication and division are replaced with approximate alternatives. For the former case, \pname-based Gaussian filter is capable of producing Superior PSNR ($24.5$), over INZeD counterpart ($20.9$). Interestingly in the latter case, not only the PSNR value of hybrid \pname~(23.3) surpasses the hybrid MBM/INZED (21.3), the values have not changed significantly compared to their "division only" approximation mode. This further motivates the deployment of hybrid \pname~mul/div.
The visual quality of two processed images are shown in Fig.~\ref{fig:gaussian}.

We also utilized our proposed SIMDive-based approximate multiplier in an ANN, provided by~\cite{mnist}, for testing its efficacy on the classification accuracy of MNIST \cite{6296535} and MNIST fashion~\cite{DBLP:journals/corr/abs-1708-07747} datasets with $28\times28$ grayscale handwritten digit images.
The ANN under consideration exploits fully connected layers with $100$ nodes for classification. Two different network configurations implemented: two 
and three fully connected hidden layers. In both configurations, the input and output layers have $784$ and $10$ neurons, respectively. For both datasets and both configurations, the network was trained with $60,000$ images using floating-point numbers; while during the inference phase for $10,000$ testing images, we quantized the network parameters and activations to 8-bit fixed point precision and evaluated the classification accuracy with accurate and approximate multipliers. The classification accuracy results are described in Table~\ref{table_ann}. Due to the inherent error resilience of ANNs, many of the quantization induced errors have been partially healed for fashion dataset. Interestingly, \pname-based ANN not only achieves same or higher classification accuracy compared to accurate and MBM/INZeD counterparts, it also outperforms accurate design and provides 22\% and 38\% improvement in terms of area and energy, respectively. 

\begin{table}[]
\centering
\caption{ANN classification accuracy with accurate/approximate Mul/Div}
  \vspace{-0.3cm}
\label{table_ann}
\def\arraystretch{1.3}
	\resizebox{\columnwidth}{!}{
\begin{tabu}{|c|c|c|c|c|c|c|}

\hline
\multirow{3}{*}{\textbf{\begin{tabular}[c]{@{}c@{}}\\MNIST \\ Dataset\end{tabular}}} & \multirow{3}{*}{\textbf{\begin{tabular}[c]{@{}c@{}}\\No. of \\ Hidden \\ layers\end{tabular}}} & \multirow{3}{*}{\textbf{\begin{tabular}[c]{@{}c@{}}\\Nodes in \\ each Hidden \\ layer\end{tabular}}} & \multicolumn{4}{c|}{\textbf{Inference Accuracy \%}} \\ \cline{4-7} 
 &  &  & \multicolumn{1}{l|}{\textbf{\begin{tabular}[c]{@{}l@{}}Double\\ Precision\end{tabular}}} & \multicolumn{3}{c|}{\textbf{8-bit Fixed Precision}} \\ \cline{4-7} 
 &  &  & \textbf{Accurate} & \textbf{Accurate} & \textbf{\pname} & \textbf{\begin{tabular}[c]{@{}c@{}}MBM/INZeD\\\cite{saadat2018minimally, saadat2019approximate}\end{tabular}} \\ \hline
\large{Digits} & \large{2} & \large{100} & \large{97.09} & \large{96.67} & \large{96.68} & \large{96.62} \\ \hline
\large{Digits} & \large{3} & \large{100} & \large{96.56} & \large{96.23} & \large{96.22} & \large{96.17} \\ \hline
\large{Fashion} & \large{2} & \large{100} & \large{85.18} & \large{84.07} & \large{84.09} & \large{84.08} \\ \hline
\large{Fashion} & \large{3} & \large{100} & \large{85.29} & \large{84.26} & \large{84.27} & \large{84.39}\\ \tabucline[1.5pt]{-------}
\multicolumn{3}{|c|}{\large{Area (normalized to 8-bit accurate)}} & - & \large{1} & \large{0.78} &\large{0.74}  \\ \hline
\multicolumn{3}{|c|}{\large{Energy (normalized to 8-bit accurate)}} & - & \large{1} & \large{0.62} & \large{0.72} \\ \hline
\end{tabu}
}
  \vspace{-0.5cm}
\end{table}
\section{Future Works and Conclusion} \label{sec:conclusion}
We proposed for the first time approximate SISD and SIMD soft multiplier-divider with better throughput and energy than the cutting-edge SIMD/SISD counterparts. In addition we proposed an accuracy control knob with our light-weight error-reduction scheme for the hybrid architecture which tunes error to a desirable bound and achieve high accuracy ($>99.2\%$).
We intend to utilize the proposed coalesced multiplier/divider in other domains, e.g. floating point units (mantissa multiplication and division). Moreover, to evaluate the potential gains of precision variability in more details, a customized SIMD-architecture of \pname-based ANN will be investigated in future tracks.
Last but not least, considering the orthogonal contribution of approximate adder/subtractor in the literature, they can be employed to add/subtract fractional part LSBs in tandem with accurate ones for MSBs without imposing high level of inaccuracy.

\bibliographystyle{ACM-Reference-Format}
\bibliography{main}

\end{document}